\documentclass[prb,twocolumn]{revtex4}
              
\usepackage{graphicx}
\usepackage{amsmath}
\usepackage{amsfonts}
\usepackage{multirow}

\begin{document}

\title{What really happens in strongly correlated superconductors:  
Insights from a quantum Monte-Carlo study of high temperature superconductivity in FeSe films
}
\author{Steven A. Kivelson}
\email{kivelson@stanford.edu}
\affiliation{Department of Physics, Stanford University, Stanford, CA 94305-4060}
\date{\today}

\maketitle

BCS theory and its extensions form the basis of a compelling and elegant understanding of the properties of ÒconventionalÓ superconductors.  However, it is a weak-coupling theory, organized around the infinitesimal ``Cooper'' instability of the Fermi liquid.  This is not merely a theoretical abstraction: Schrieffer (among others) has  emphasized that it is only because the ÒnormalÓ state above $T_c$ is so accurately described by Fermi liquid theory that we can talk about  collective pairing of well-defined quasiparticles. Moreover, the fact that the phonons involved are coherent collective modes  permits the identification of a pairing Òglue,Ó while the small ratio of the phonon frequencies to the Fermi energy  allows the relatively weak ÒretardedÓ attraction they induce to overcome the strong Coulomb repulsion between electrons.  

When it comes to  high temperature superconductors, all the features of the problem that guarantee the successes of BCS theory are missing:  There are no well defined quasiparticles in the normal state at $T>T_c$; any collective modes that have been proposed as candidate replacements for the phonons ({\it e.g.} antiferromagnetic spin fluctuations) are overdamped excitations which are only coherent in a putative ``parent'' state; there is little dynamic range for retardation.  Indeed, the repulsive Coulomb interactions, themselves, surely play a key role in driving the pairing.  Loosely, this is what is meant when the relevant materials are said to be ``highly correlated.''  In general, analytic approaches to this problem are forced to adopt an essentially weak-coupling perspective and to rely on an assumed smooth adiabatic connection that allows weak-coupling results to be extrapolated to the  intermediate coupling regime.

An alternative theoretical approach is to attempt a numerical solution of simple lattice-scale models of interacting electrons.  The only method that currently appears capable of yielding solutions of such problems in more than one dimension and for large enough size systems to be useful for extracting reliable phase diagrams is one or another form of quantum Monte Carlo.  For generic problems involving electrons, this is not possible due to the infamous ``fermion sign problem.'' However, for various classes of models\cite{Li_2016,Wu_2005,Li_2015}, sign-problem-free determinental quantum Monte-Carlo (DQMC) studies are possible.  Significantly, it has recently been realized\cite{Li_2016,Berg_2012,Schattner_2015,Li_2016b,Dumitrescu_2016,Schattner_2016} that this solvable class includes models that are plausibly relevant to the physics of high temperature superconductivity.  

In Ref. [1], Li {\it et al} have introduced an explicit model that captures  salient low-energy features of the electronic structure of single-layer FeSe on STO, one of the most interesting high temperature superconductors.\cite{FeSe} Specifically, the model has a bandstructure with the same pair of electron pockets at the Fermi energy that have been observed in ARPES experiments.  Moreover, without sacrificing the conditions necessary for the model to be solvable by DQMC, the effects of near quantum-critical antiferromagnet fluctuations of various sorts, as well as coupling to optic phonons and to nematic fluctuations, can be explored in the context of this model.  Of course, in addition to the practical problem that such calculations are extremely time-consuming and demanding of computational resources, the downsides of such an approach are that one would always like to be able to study still larger systems (the largest systems studied by Li {\it et al} are $22 \times 22$), that dynamical quantities are difficult to extract, and that the results are purely numerical so that it is extremely difficult to determine what effects would result from the introduction of this or that perturbation.  However, the benefit of this approach is that the answers are not dependent on the accuracy of any a priori guess concerning the importance of particular physical processes, nor are they in any sense perturbative.  For studying the properties of highly correlated materials, this is an extremely valuable feature.

Single layer FeSe is a particularly interesting example of a high temperature superconductor for several reasons.  Firstly, it has the highest superconducting $T_c$ observed to date among the Fe based superconductors.  This makes it critical to address the question of what is special about this material.  Secondly, its Fermi surface is relatively simple -- it consists of two symmetry related  electron pockets.  This is in contrast with the usual case in the Fe based superconductors where there are both electron and hole pockets.  Whatever the role this simple fermiology plays in the physics of $T_c$, from the technical viewpoint it is a fortuitous feature of the problem;  the effects of  inter-pocket scattering by magnetic fluctuations can be studied without  minus-sign problems.  

In the literature concerning the mechanism of superconductivity in the Fe based superconductors, prominent roles have been suggested for various forms of collective fluctuations including those associated with putative proximate phases with antiferromagnetic, nematic, and/or various forms of orbital order.  Given that lattice vibrations are a universal feature of all solids, it is also reasonable to consider the role of optical phonons, as in conventional superconductors.  Because high temperature superconductivity is ubiquitous in the Fe-pnictide/Fe-chalcoginide family, Li {\it et al} (quite reasonably) adopted the perspective that the principle interaction responsible for superconductivity is unlikely to be specific to FeSe films. Thus, they set about to identify a ``principle'' interaction, and a ``secondary'' (or helper) interaction which can boost $T_c$ -- the latter possibly associated with a specific feature of the FeSe films.

There are many reasons (which have been discussed in many places) to believe that short-range correlated antiferromagnetic fluctuations are the prime drivers of the superconducting pairing in many -- possibly all -- high temperature superconductors.  Where  magnetic order is observed in  Fe-based superconductors, it is ususally  ``stripe-like,'' with an ordering vector that is nearly ideal for scattering electrons between the electron and hole pockets.  From this perspective, the fact that the FeSe films  have the highest $T_c$'s while lacking the requisite hole pockets is a challenge for theories of magnetic pairing.  The work of Li {\it et al} does not directly address this, as the magnetic coupling studied is peaked at the ``N{\'e}el'' wave vector, $(\pi,\pi)$, which allows scattering between the two electron pockets.  However, the present results do show that various types of magnetic fluctuations are able to mediate strong superconducting pairing, and that this effect persists far into the magnetically disordered phase where the magnetic correlations are presumably quite short-ranged. 

{\em Perhaps the most striking new result in the present study concerns the effectiveness of helper fluctuations.}  Addition of either electron-phonon scattering or nearly quantum critical nematic fluctuations results in a factor of 2 or more  enhancement of the pairing amplitude.

In this context, an issue of direct physical interest is whether it is possible to get a substantial enhancement of superconductivity by coupling to an optical phonon with a coupling constant that is strong only very near $\vec q=\vec 0$.  One of the most striking features revealed in 
experimental studies of 
single layer FeSe on STO is the presence of a mysterious ``replica band'' below the Fermi energy.  A possible interpretation\cite{Lee_2014} of this band is that it reflects a shake-off process involving an optical phonon with a sharply defined energy {\em and momentum.}  This requires an unusual form of the electron-phonon coupling that is concentrated in a range of momenta that is small compared to $2k_F$.   
The relevant coupling for superconductivity is integrated over the entire Fermi surface.  Thus, in general, 
 if the electron phonon coupling were  substantial only for a range of momentum transfers $\Delta q \ll 2k_F$, the effectiveness of the coupling for enhancing superconductivity would be reduced parametrically in proportional to $\Delta q/2k_F$.  Precisely this issue arose in the context of the role of the ``resonant mode'' in 
  the cuprates\cite{Kee_2002};  the resonant mode is the most dramatic and sharply defined feature in the magnetic excitation spectrum of the superconducting cuprates, but because it is confined to a small portion of the Brillouin zone, its integrated weight is relatively small.  Thus it probably plays little direct role in mediating pairing interactions.  

In the model of electron-phonon coupling considered by Li {\it et al} 
 the range of momentum transfer involved is determined by the size of the Fermi pockets, {\it i.e.} $2k_F$.  Thus, to test this issue, the model treated would need to be modified so that the electron-phonon coupling itself was strongly momentum dependent.  To get a strong effect from a small range of momenta, the strength of the coupling at $\vec q=\vec 0$ must be anomalously strong  to compensate for the  suppression by $\Delta q/2k_F$. In the case of nematic criticality, the coupling at $\vec q=\vec 0$ is proportional to the nematic susceptibility\cite{Lederer_2015}, which diverges at the nematic QCP.  In the case of an optically active phonon in STO, what is needed is a large ``dynamical charge,'' $eZ^{eff}$, where  the coupling at $\vec q=\vec 0$ is proportional to $|Z^{eff}|^2$.  Large dynamical charges are uncommon, but can occur in nearly ferroelectric oxides such as STO.\cite{STO}

Going forward, this paper is an avatar of a new prong in the theory of high temperature superconductivity Ð and of highly correlated systems more generally.  The numerical studies of Li {\it et al } are impressive and their key conclusions are important and seem strong.  However, in common with all numerical studies, there are issues of interpretation that are worthy of further testing and investigation.  Most obviously, the specific Monte Carlo method employed is a $T=0$ method -- thus, $T_c$ is never directly measured, but is rather inferred from a (plausible) proportionality between $T_c$ and the (measured) superconducting gap.  
 At a more picky level, the existence of a superconducting state is inferred from a finite size extrapolation to the thermodynamic limit of the long-distance value of the superconducting pair-field correlation function.  At the system sizes accessible to this study, this correlation function is still strongly size dependent, so this extrapolation is not entirely unambiguous.  Of course, studies on larger size systems would be useful, but perhaps still more direct\cite{Schattner_2015} would be measures of the finite $T$ superfluid stiffness, which in 2d is the most direct measure of  the superconducting state.  

\end{document}